\documentclass[conference]{IEEEtran}
\IEEEoverridecommandlockouts

\usepackage{algorithm}
\usepackage{algpseudocode}

\usepackage{graphicx}
\graphicspath{ {./} }
\usepackage[english]{babel}
\usepackage{blindtext}
\usepackage[inline]{enumitem}
\usepackage{enumitem}
\usepackage{cite}
\usepackage{balance}
\usepackage{amsmath,amssymb,amsfonts}
\usepackage{subcaption}
\usepackage[font=footnotesize]{caption}
\usepackage{textcomp}
\usepackage{xcolor}

\newtheorem{lemma}{Lemma}

\def\BibTeX{{\rm B\kern-.05em{\sc i\kern-.025em b}\kern-.08em
		T\kern-.1667em\lower.7ex\hbox{E}\kern-.125emX}}

\newcommand{\norm}[1]{  \left\Vert  #1 \right\Vert}

\newcommand{\her}{\mathsf{H}}

\newcommand{\trp}{{\mathsf{T}}}

\usepackage{setspace}

\begin{document}
	
	\title{Digital Twin Aided RIS Communication: \\ Robust Beamforming and Interference Management}
	\author{\IEEEauthorblockN{Sadjad Alikhani and Ahmed Alkhateeb}
		\IEEEauthorblockA{Wireless Intelligence Lab, Arizona State University, USA}{Emails: \{alikhani, alkhateeb\}@asu.edu}}
	\maketitle
	
	\begin{abstract}
		Reconfigurable intelligent surfaces (RISs) are envisioned to play a key role in future wireless communication networks. However, channel estimation in RIS-aided wireless networks is challenging due to their passive nature and the large number of reflective elements, leading to high channel estimation overhead. Additionally, conventional methods like beam sweeping, which do not rely on explicit channel state information, often struggle in managing interference in multi-user networks. In this paper, we propose a novel approach that leverages \textit{digital twins} (DTs) of the physical environments to approximate channels using electromagnetic 3D models and ray tracing, thus relaxing the need for channel estimation and extensive over-the-air computations in RIS-aided wireless networks. To address the digital twins channel approximation errors, we further refine this approach with a DT-specific robust transmission design that reliably meets minimum desired rates. The results show that our method secures these rates over \(90\%\) of the time, significantly outperforming beam sweeping, which achieves these rates less than \(8\%\) of the time due to its poor management of transmitting power and interference. 
	\end{abstract}
	\section{Introduction}
	Reconfigurable intelligent surfaces (RISs) are envisioned as integral components of beyond-5G wireless systems thanks to their ability to dynamically control the electromagnetic waves. These programmable metasurfaces offer precise control over signal propagation. However, channel estimation in RIS-aided networks poses a significant challenge due to the passive nature of RIS elements, which lack RF chains. This makes direct estimation of channels between the base station (BS) and RIS or RIS and users difficult. Instead, cascaded channel estimation is employed, a process that grows more complex with the increase in the number of RIS elements \cite{9732214}. Overcoming these channel estimation difficulties is crucial for harnessing RIS's potential to enhance spectral and energy efficiencies and expand network coverage.
	\par \textbf{Prior Work:} 
	The RIS channel estimation approaches in the literature mostly result in high pilot training overhead, insufficient adaptability to millimeter-wave (mmWave) networks, high power leakage, significant estimation errors, or relying on impractical assumptions \cite{9053695,9130088,9103231,chen2023channel,9732214}. To address these challenges, considerable research has shifted focus to beam sweeping, which aims to find the predefined beams that yield the highest performance through multiple interactions with users, without relying on explicit channel state information (CSI) \cite{10057262,tian2021fast,singh2021fast,9539048}. However, most of this research is confined to single-user networks. In multi-user RIS-aided networks, beam sweeping often results in extensive over-the-air computations and interactions with users and needs very large codebooks for interference management. Additionally, relying solely on user positions knowledge is ineffective as it ignores fading and multipath effects, where obstructions can severely impact the quality of service.
	\par \textbf{Contribution:}
	In this paper, we explore a novel direction where digital twins of the physical environments are utilized to address channel estimation challenges in RIS-aided networks. These twins are pivotal in simulating and optimizing complex systems and particularly effective at accurately approximating wireless channels. By creating virtual replicas of real-world wireless environments, digital twins enable the exploration of channel characteristics, system optimization, and the development of advanced signal processing techniques \cite{10198573}. More specifically, in this work, to tackle the challenge of estimating channels in RIS-aided wireless networks, and inspired by recent advancements in real-time digital twins \cite{10198573}, we introduce a novel digital twin aided framework, with the following key contributions:
	\begin{itemize}
		\item We develop a novel digital twin-based beamforming and phase-shift design in RIS-aided wireless networks that approximates channels using electromagnetic (EM) 3D models and ray-tracing, rather than relying on explicit CSI.
		\item We enhance the proposed DT-based approach with a robust transmission design that reliably meets minimum desired rates, even in scenarios where the DT channels may include approximation errors.
		\item Additionally, we incorporate three novel considerations into our robust transmission design, establishing a foundation for more practical designs.
	\end{itemize}
	\begin{figure*} [t]
		\centerline{\includegraphics[width=\textwidth]{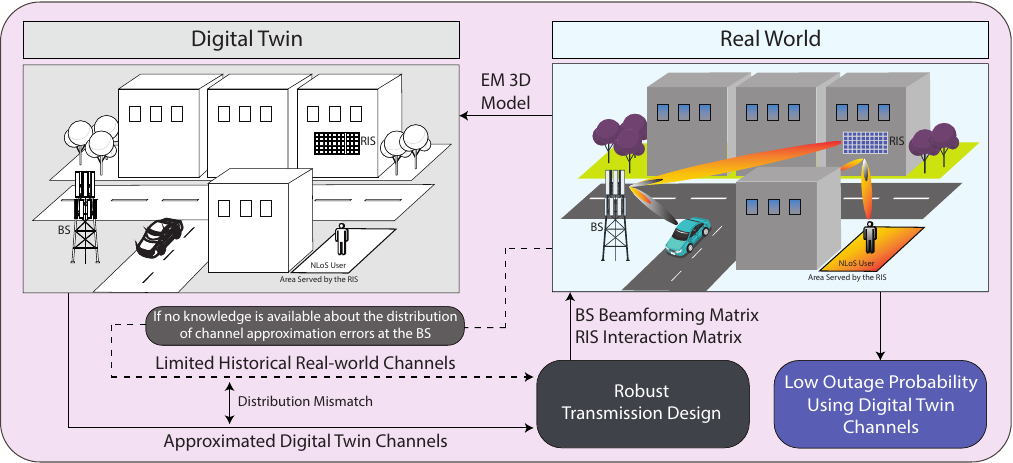}}
		\caption{This figure illustrates our system model, depicting a scenario where the line-of-sight user is exclusively served by the base station, while the non-line-of-sight user is served solely by a reconfigurable intelligent surface. Since channel estimation approaches in RIS-aided wireless networks are typically infeasible for real-time applications, we use digital twins of the physical environments to approximate the channels. Additionally, we develop a robust transmission design that accounts for digital twin channel approximation errors and reliably meets the network requirements.
		}
		\label{fig:SystemModel}
	\end{figure*}
	\section{System and Signal Model} \label{sec:sysModel}
	We consider an RIS-aided wireless communication setup as shown in Fig. \ref{fig:SystemModel}. The BS is equipped with $N_t$ antennas, the RIS comprising $M$ elements, and $K=2$ users each equipped with $N_r$ antennas. One of the two users has no direct path to the BS and can only be served by the RIS, while the other user is exclusively served by the BS without any assistance from the RIS. For each user \( k \in \left[K\right]\), define \(\mathbf{H}_k \in \mathbb{C}^{N_r \times N_t}\) as the downlink channel from the BS to user \( k \), \(\mathbf{G}_k \in \mathbb{C}^{N_r \times M}\) as the channel from the RIS to user \( k \), and \(\mathbf{H}_{BR} \in \mathbb{C}^{M \times N_t}\) as the channel from the BS to the RIS. Then, in the downlink, the received signal at user $k$ is given by
	\begin{subequations}
		\begin{align}
			& \mathbf{y}_k = \left( \mathbf{H}_k + \mathbf{G}_k \mathbf{\Theta} \mathbf{H}_{BR}\right)  \sum_{k=1}^{K} \mathbf{f}_k x_k + \mathbf{n}_k,
		\end{align}
	\end{subequations}
	where $\mathbf{\Theta} = \mathrm{diag}(\mathbf{\theta}) \in \mathbb{C}^{M \times M}$ is the diagonal matrix containing the RIS phase shifts on its diagonal. The beamforming vector for user $k$ at the BS is $\mathbf{f}_k \in \mathbb{C}^{N_t}$, and $x_k$ is the signal transmitted to user $k$, with the expected power $\mathbb{E}[\lvert x_k \rvert^2] = 1$. The noise $\mathbf{n}_k$ at user $k$ is complex additive white Gaussian noise with zero mean and variance $\sigma_k^2$, i.e., $\mathbf{n}_k \sim \mathcal{CN}(0, \sigma_k^2 \mathbf{I})$.
	\section{Problem Formulation}
	The considered RIS-aided communication scenario has two important aspects: It has multiple users and, generally, each user could be served by both the BS and the RIS. Our objective in this paper is to optimize the RIS reflection matrix $\mathbf{\Theta}$ and BS beamforming to ensure that each user at least achieves the minimum spectral efficiency (SE). Given the system model in section \ref{sec:sysModel}, the SE for each user \( k \) can be written as
	\begin{subequations}
		\begin{align}	 
			& \mathsf{SE}_k = \log \left(1 + \frac{\lVert  \left( \mathbf{H}_k + \mathbf{G}_k \mathbf{\Theta} \mathbf{H}_{BR}\right) \mathbf{f}_k \rVert^2_2}{ \sum_{u \neq k}\lVert  \left( \mathbf{H}_k + \mathbf{G}_k \mathbf{\Theta} \mathbf{H}_{BR}\right) \mathbf{f}_u \rVert^2_2 + \sigma^2_k}\right).
		\end{align}
	\end{subequations}
	Different from the max-min fairness problem that aims to equalize the distribution of resources as much as possible by maximizing the minimum rate among all users \cite{4346554}, our focus here is on meeting predefined SEs for each user with minimum total transmitting power at the BS \cite{9180053,9650619,9322565}. As a result, we formulate an optimization problem as follows 
	\begin{subequations} \label{eq:sumrate}
		\begin{align}
			&\min_{\mathbf{F},\mathbf{\Theta}} \hspace{.4cm}  \lVert \mathbf{F} \rVert^2_\mathsf{F} \\
			&\hspace{.25cm} \text{\normalfont s.t.} \hspace{.5cm} \mathsf{SE}_k \geq \gamma_k, \hspace{.25cm} k \in \left[K\right] \label{eq:se}\\
			& \hspace{1.15cm} \lvert \mathbf{\theta}_{m} \rvert = 1, \hspace{.25cm} m \in \left[M\right]  \label{eq:unit-mod},
		\end{align}
	\end{subequations}
	where \( \mathbf{F} = \left[\mathbf{f}_1, \mathbf{f}_2, \cdots , \mathbf{f}_K \right] \) and \( \gamma_k \) is the minimum SE target for user \( k \). This optimization problem, however, is non-convex because
	\begin{enumerate*}[label=(\roman*)]
		\item the optimization variables, which are beamforming vectors and phase shifts, are coupled,
		\item the constraint (\ref{eq:se}) involves a logarithmic ratio which is non-convex, and
		\item the unit-modulus constraint (\ref{eq:unit-mod}) on phase shifts is a non-convex constraint.
	\end{enumerate*} 
	What makes this problem challenging is solving it in real-world scenarios. To relax this problem, we will assume that one user is only served by the BS and the other user, which has no direct path to the BS, is solely served by the RIS. With that, problem (\ref{eq:sumrate}) can be simplified as
	\begin{subequations} \label{eq:decoupledOpt}
		\begin{align} 
			&\min_{\mathbf{f}_1,\mathbf{f}_2,\mathbf{\Theta},\mathbf{\varepsilon}_1,\mathbf{\varepsilon}_2} \hspace{.1cm}\lVert \mathbf{f}_1 \rVert^2_2 + \lVert \mathbf{f}_2 \rVert^2_2\\
			&\hspace{.8cm} \text{\normalfont s.t.} \hspace{.5cm} \lVert \mathbf{G}_2 \mathbf{\Theta} \mathbf{H}_{BR} \mathbf{f}_1 \rVert^2_2 + \sigma_2^2 \leq \varepsilon_2, \label{eq:constr1}\\ 
			& \hspace{1.7cm} \lVert  \mathbf{G}_2 \mathbf{\Theta} \mathbf{H}_{BR}\mathbf{f}_2 \rVert^2_2 \geq \varepsilon_2 \left( 2^{\gamma_2} - 1\right), \label{eq:constr2}\\
			& \hspace{1.7cm}  \lVert  \left( \mathbf{H}_1 + \mathbf{G}_1 \mathbf{\Theta} \mathbf{H}_{BR} \right)\mathbf{f}_2 \rVert^2_2 + \sigma_1^2  \leq \varepsilon_1, \label{eq:constr3}\\ 
			&\hspace{1.7cm} \lVert   \mathbf{H}_1 \mathbf{f}_1 \rVert^2_2 \geq \varepsilon_1 \left( 2^{\gamma_1} - 1\right),\label{eq:constr4}\\ 
			& \hspace{1.7cm} \lvert \mathbf{\theta}_m \rvert = 1, \hspace{.25cm} m \in \left[M\right] \label{eq:unitMod}.
		\end{align}
	\end{subequations}
	This revised problem remains non-convex but is less complex to solve. Solving (\ref{eq:decoupledOpt}) in real systems, however, still has a \textbf{significant challenge as it assumes the availability of global channel knowledge at both the BS and the RIS}. Further, the lack of RF chains in passive RISs makes it almost impossible to estimate its channels, without relying on cascaded channel estimation approaches that require massive training overhead. We will next introduce our proposed DT-based solution to address this challenge.
	\section{Proposed Digital Twin Based Solution}
	To meet each user's spectral efficiency requirements and minimize transmitting power at the BS, it is essential to solve the optimization problem (\ref{eq:decoupledOpt}). However, accurately meeting these quality of service (QoS) constraints is difficult without the knowledge of cascaded channels in RIS-aided communications. Additionally, traditional methods like beam sweeping that do not rely on explicit CSI are often ineffective in multi-user scenarios. We address these challenges using digital twins to approximate these channels and effectively solve (\ref{eq:decoupledOpt}). We will discuss the underlying concepts and our adopted digital twin model in the subsequent subsections.
	\subsection{Key Idea}
	We propose a framework where the BS uses digital twins of the physical environment to enhance network performance, outlined with two main components \cite{10198573}:
	\begin{enumerate*}[label=(\roman*)]
		\item EM 3D Model Extractor: Generates a detailed 3D model capturing the electromagnetic properties necessary for precise simulations.
		\item Ray-Tracing Unit: Simulates radio wave propagation using the EM 3D model, providing a realistic representation of wireless channels.
	\end{enumerate*}
	These tools enable the approximation of wireless channel conditions without resource-intensive channel estimation and enhance the ability of BS to efficiently manage power and ensure robust connections by solving (\ref{eq:decoupledOpt}).
	\subsection{Adopted Digital Twin Model}
	In the adopted digital twin model, the real-world communication channel is influenced by key factors \cite{jiang2023digital}:
	\begin{enumerate*}[label=(\roman*)]
		\item Environmental Factors: Include the positions, movements, shapes, and materials of base stations, user equipment, and other elements like reflectors and scatterers, collectively noted as \(\mathcal{E}\).
		\item Wireless Signal Propagation Law: Represented by \(g(\cdot)\), this dictates signal propagation through the environment, affected by the environmental factors.
	\end{enumerate*}
	The channel is modeled without direct estimation by \(\mathbf{H} = g(\mathcal{E})\). Digital twins enhance this with EM 3D models (\(\widetilde{\mathcal{E}}\)) and ray tracing (\(\tilde{g}(\cdot)\)), approximating the channel as \(\widetilde{\mathbf{H}} = \tilde{g}(\widetilde{\mathcal{E}})\). This simulation method provides detailed insights into channel behaviors, aiding in efficient network management and optimization.
	\section{Beamforming and RIS Interaction Optimization with Perfect Digital Twin Channels}
	With digital twins providing approximations of all necessary channels, we are set to tackle the optimization problem (\ref{eq:decoupledOpt}) based on our simplified yet practical system model. While digital twins offer a simplified channel approximation, potential inaccuracies underscore the need for careful consideration during the network optimization process. Initially, we will solve this problem assuming perfect digital twin channels for a best-case scenario, and later develop a robust transmission design to accommodate possible errors in digital twin channel approximations, aiming to consistently meet minimum desired rates. In scenarios with perfect digital twin channels, beamforming and RIS interactions are designed by solving (\ref{eq:decoupledOpt}). This problem is inherently non-convex due to
	\begin{enumerate*}[label=(\roman*)]
		\item coupled beamformers and phase-shift matrix,
		\item quadratic convex functions greater than constants in constraints, and
		\item unit-modulus phase-shifts constraint.
	\end{enumerate*}
	To make this problem convex, we utilize alternating optimization (AO), affine transformation, and semi-definite relaxation (SDR) approaches. The affine transformation allows the general term in (\ref{eq:constr1}-\ref{eq:constr4}) to be recast as
	\begin{equation} \label{eq:affine}
		\lVert \left(\mathbf{D} + \mathbf{G} \mathbf{\Theta} \mathbf{H} \right) \mathbf{f} \rVert^2_2 = \sum_{r=1}^{R} \mathrm{tr} \left( \mathbf{\Gamma}_{1,r} \mathbf{F} \right) = \sum_{r=1}^{R} \mathrm{tr} \left( \mathbf{\Gamma}_{2,r} \mathbf{E} \right),
	\end{equation}
	where $\mathbf{\Gamma}_{1,r} = \mathbf{\Upsilon}_r^\her \mathbf{E} \mathbf{\Upsilon}_r$, $\mathbf{\Gamma}_{2,r} = \mathbf{\Upsilon}_r \mathbf{F} \mathbf{\Upsilon}_r^\her$, $ \mathbf{\Upsilon}_r = \begin{bmatrix} \mathbf{D}_r \\  \mathbf{F}_r \end{bmatrix}$, $\mathbf{E} = \mathbf{\theta}^{\prime,\her} \mathbf{\theta}$, $\mathbf{\theta}^\prime = \begin{bmatrix} 1 & \mathbf{\theta} \end{bmatrix}$, $\mathbf{F} = \mathbf{f} \mathbf{f}^\her$, $\mathbf{F}_r = \mathrm{diag} \left(\mathbf{G}_r\right) \mathbf{H}$, and $\mathbf{G}_r$ and $\mathbf{D}_r$ are the $r$-th row of $\mathbf{G}$ and $\mathbf{D}$ with $R$ rows, respectively. By applying (\ref{eq:affine}) to constraints (\ref{eq:constr1}-\ref{eq:constr4}), we establish convex constraints. Despite the non-convex rank-one constraints introduced by the affine transformation, they are manageable via the SDR method combined with Gaussian randomization. Employing SDR, the constraint (\ref{eq:unitMod}) is relaxed to allow feasible set mapping with each AO iteration. These adaptations convert (\ref{eq:decoupledOpt}) into a convex problem solvable by the CVX tool\cite{cvx}.
	\section{Digital Twin-Based Robust Transmission Design}
	Given the potential for errors in digital twin approximations, it is essential to ensure our solutions meet the minimum desired rates for all users. These errors can arise from inaccuracies in 3D models, ray-tracing, and user localization. Although the channel between the BS and the RIS is typically precise, the channels from the BS to the first user (\(\mathbf{H}_1\)), from the RIS to the first user (\(\mathbf{G}_1\)), and from the RIS to the second user (\(\mathbf{G}_2\)) are particularly prone to errors. To address these discrepancies, we introduce a channel error model for the \(r\)-th antenna of users as follows
	\begin{subequations} \label{eq:errModel}
		\begin{align}
			\mathbf{\Upsilon} &= \widetilde{\mathbf{\Upsilon}} + \Delta {\mathbf{\Upsilon}} \nonumber \\
			&= \begin{bmatrix} \widetilde{\mathbf{D}} \\ \mathrm{diag} \left(\widetilde{\mathbf{G}} \right) \mathbf{H} \end{bmatrix} + 
			\begin{bmatrix} \Delta{\mathbf{D}} \\ \mathrm{diag} \left(\Delta{\mathbf{G}} \right) \mathbf{H} \end{bmatrix},
		\end{align}
	\end{subequations}
	where \(\widetilde{\mathbf{B}}\) and \(\Delta{\mathbf{B}}\) denote the digital twin approximation of \(\mathbf{B}\) and its error, respectively. This model defines an accurate channel as the sum of its approximation and the error. This strategy enables robust transmission solutions that consistently satisfy the QoS requirements for all users. We will next outline our robust transmission design and proceed with solving the associated optimization problem.
	\subsection{Errors Specific to Digital Twins} 
	To tackle errors in practical digital twins, we propose three strategies for robust transmissions:
	\begin{enumerate*}[label=(\roman*)]
		\item DT-Specific Error Models: We create error models tailored for channels approximated by digital twins, distinct from traditional estimation models.
		\item Flexible Design: Our approach accommodates any type of channel error covariance matrix, enhancing adaptability.
		\item Error Learning: We devise a method to learn channel approximation error characteristics without prior knowledge, moving beyond conventional assumptions.
	\end{enumerate*}
	These strategies address key inaccuracies from ray-tracing and EM 3D models, focusing on differences between DT channels and real-world channels by analyzing their strongest paths. We will delve into these approaches further.
	\par \textbf{Characteristics of DT Channel Approximation Errors:} Our numerical results show that the approximation errors from digital twins, which do not fully capture all environmental reflections and scatterings, generally adhere to a distribution very close to a complex Gaussian distribution.
	\par \textbf{Distribution of DT Channel Approximation Errors:} Contrary to common assumptions in the literature that assume identically distributed channel error columns, typically modeled as \( \mathrm{vec} \left( \Delta {\mathbf{\Upsilon}} \right) \sim \mathcal{CN} \left( \mathbf{0}, \varepsilon \mathit{\mathbf{I}} \right)\) \cite{9322565,9650619,9180053},
	our results suggest a more complex structure. We introduce a more general model for robust transmission design to handle diverse channel error covariance matrices, represented by
	\begin{equation} \label{eq:distrErr}
		\mathrm{vec} \left( \Delta {\mathbf{\Upsilon}} \right) \sim \mathcal{CN} \left( \mathbf{0}, \mathbf{\Sigma} \right),
	\end{equation}
	with the covariance matrix \(\mathbf{\Sigma}\) structured as
	\begin{equation}
		\mathbf{\Sigma} = \begin{bmatrix} \mathbf{\Sigma}_{D} & \mathbf{\Sigma}_{D,F} \\ \mathbf{\Sigma}_{F,D} & \mathbf{\Sigma}_{F} \end{bmatrix}.
	\end{equation}
	Here, despite inaccuracies due to ray-tracing/EM 3D model limitations, the strongest paths captured by the DT channels essentially reflect the core attributes of the actual channels. Subsequent weaker paths in the real-world channels make minor adjustments that typically corroborate the approximations made by the strongest paths in digital twins. Given this alignment, and to simplify our calculations, we treat the mean of these approximation errors as zero.
	\par \textbf{DT-based Channel Error Learning:} 
	Prior work on robust transmission design often assumes that the BS knows the error covariance matrix, typically \(\mathbf{\Sigma} = \varepsilon \mathit{\mathbf{I}}\), but does not specify how this knowledge is acquired. We introduce a DT-based method that empirically determines these parameters. In this method, during the initial coherence blocks in a specific zone, where channel errors are similar, the digital twin utilizes precise historical real-world channel data from conventional methods. This allows the digital twin to empirically establish the mean and covariance of channel errors. Over time, this approach reduces the need for expensive estimated channels, enabling the BS to develop robust transmissions more efficiently and practically. The sample covariance matrix at the \(n\)-th channel coherence block is calculated as follows
	\begin{align} \label{eq:empCov}
		\Sigma^{\left(n\right)} = &\left(1-\frac{1}{n}\right) \Sigma^{\left(n-1\right)} \nonumber \\
		& + \frac{1}{n} \mathrm{vec} \left(\Delta\mathbf{\Upsilon}^{\left(n\right),\her}\right) \mathrm{vec}^\her \left(\Delta\mathbf{\Upsilon}^{\left(n\right),\her}\right).
	\end{align}
	\subsection{Robust Transmission Design}
	Our robust transmission approach focuses on optimizing the probability of constraint satisfaction in (\ref{eq:decoupledOpt}). By defining \(\rho\) as the outage probability, we reformulate this consideration as follows
	\begin{equation} \label{eq:prob}
		\mathrm{Pr}\{\text{Constraint Satisfaction}\} \geq 1-\rho.
	\end{equation}
	This formulation allows us to convert probabilistic constraints into deterministic ones through the application of the following lemma \cite{9180053}.
	\begin{lemma} \label{lemma:bern}
		(Bernstein-Type Inequality) Let \(f\left(\mathbf{x}\right) = \mathbf{x}^\her \mathbf{U} \mathbf{x} + 2 \mathrm{Re} \left\lbrace \mathbf{u}^\her \mathbf{x} \right\rbrace + u\), where \(\mathbf{U} \in \mathbb{H}^{n \times n}\), \(\mathbf{u} \in \mathbb{C}^{n \times 1}\), \(u \in \mathbb{R}\), and \(\mathbf{x} \in \mathbb{C}^{n \times 1}  \sim \mathcal{CN} \left( \mathbf{0},\mathit{\mathbf{I}} \right)\). Then for any \(\rho \in [0,1]\), the following approximation is valid:
		\begin{subequations}
			\begin{align}
				&\mathrm{Pr} \left\lbrace   \mathbf{x}^\her \mathbf{U} \mathbf{x} + 2 \mathrm{Re} \left\lbrace \mathbf{u}^\her \mathbf{x} \right\rbrace + u \geq 0 \right\rbrace \geq 1 - \rho \\
				&\Rightarrow \mathrm{tr} \left\lbrace \mathbf{U} \right\rbrace - \sqrt{2 \mathrm{ln} \left( 1 / \rho \right)} x + \mathrm{ln} \left( \rho \right) \lambda_{\mathrm{max}}^+ \left( -\mathbf{U} \right) + u \geq 0 \\
				&\Rightarrow \begin{cases}
					\mathrm{tr}\left\lbrace \mathbf{U} \right\rbrace - \sqrt{2 \mathrm{ln} \left( 1 / \rho \right)} x + \mathrm{ln} \left( \rho \right) y + u \geq 0 \\
					\sqrt{\norm{\mathbf{U}}_\mathsf{F}^2 + 2\norm{\mathbf{u}}^2} \leq x \\
					y \mathit{\mathbf{I}} + \mathbf{U} \succeq \mathbf{0} , y \geq 0, 
				\end{cases} \label{eq:bernConstrs}
			\end{align}
		\end{subequations}
		where \(\lambda_{\mathrm{max}}^+ \left( -\mathbf{U} \right) = \mathrm{max} \left( \lambda_{\mathrm{max}} \left( -\mathbf{U} \right), 0 \right)\). Here, \(\mathbf{x}\) and \( \mathbf{y}\) are used as slack variables.
	\end{lemma}
	\par To align the form of constraints (\ref{eq:constr1}-\ref{eq:constr4}) with the setup required by Lemma \ref{lemma:bern}, we incorporate the error model (\ref{eq:errModel}) for a suitable reformulation. We can proceed to modify the general term $\lVert  \left( \mathbf{D} + \mathbf{G} \mathbf{\Theta} \mathbf{H} \right) \mathbf{f} \rVert^2_2$ in (\ref{eq:affine}) as follows
	\begin{align} \label{eq:csiErrModelInSimplify}
		& \lVert  \left( \mathbf{D} + \mathbf{G} \mathbf{\Theta} \mathbf{H} \right) \mathbf{f} \rVert^2_2 = \sum_{r=1}^{R} \lVert \left( \mathbf{D}_r + \mathbf{G}_{r} \mathbf{\Theta}  \mathbf{H} \right) \mathbf{f} \rVert^2_2  \nonumber \\
		& = \sum_{r=1}^{R} \lVert  \mathbf{\theta}^\prime  \mathbf{\Upsilon}_r \mathbf{f} \rVert^2_2   = \sum_{r=1}^{R} \lVert  \mathbf{\theta}^\prime \left(\widetilde{\mathbf{\Upsilon}}_r + \Delta {\mathbf{\Upsilon}_r}\right) \mathbf{f} \rVert^2_2.
	\end{align}
	Each term in (\ref{eq:csiErrModelInSimplify}) can primarily be managed by applying the mathematical transformations $\mathrm{tr}\left(\mathbf{X}_1 \mathbf{X}_2\right) = \mathrm{tr}\left(\mathbf{X}_2 \mathbf{X}_1\right)$, $\mathrm{tr}\left(\mathbf{X}_1^\her \mathbf{X}_2\right) = \mathrm{vec}^\her\left(\mathbf{X}_1\right)\mathrm{vec}\left(\mathbf{X}_2\right)$, and $\mathrm{tr}\left(\mathbf{X}_1 \mathbf{X}_2 \mathbf{X}_{3} \mathbf{X}_4^\her\right) = \mathrm{vec}^\her\left(\mathbf{X}_4\right)\left(\mathbf{X}_{3}^\trp \otimes \mathbf{X}_1\right) \mathrm{vec}\left(\mathbf{X}_2\right)$  as follows
	\begin{align} \label{eq:formulateToBern}
		&\lVert  \mathbf{\theta}^\prime \left(\widetilde{\mathbf{\Upsilon}} + \Delta {\mathbf{\Upsilon}}\right) \mathbf{f} \rVert^2_2 \nonumber \\
		&= \mathrm{tr} \left(\mathbf{E} \Delta {\mathbf{\Upsilon}}_n \mathbf{F} \Delta {\mathbf{\Upsilon}}^\her \right) 
		+ 2\mathrm{Re} \left\{\mathrm{tr} \left(\mathbf{E} \Delta {\mathbf{\Upsilon}} \mathbf{F} \widetilde{\mathbf{\Upsilon}}^\her \right)\right\} \nonumber \\
		& \hspace{.4cm} + \mathrm{tr} \left(\mathbf{E} \widetilde{\mathbf{\Upsilon}} \mathbf{F} \widetilde{\mathbf{\Upsilon}}^\her \right) \nonumber \\
		& = \mathrm{vec}^\her \left(\Delta {\mathbf{\Upsilon}}^\her\right) \left(\mathbf{E}^\trp \otimes \mathbf{F} \right) \mathrm{vec} \left(\Delta {\mathbf{\Upsilon}}^\her\right) \nonumber \\
		& \hspace{.4cm} + 2\mathrm{Re} \left\{\mathrm{vec}^\her \left(\mathbf{F}  \widetilde{\mathbf{\Upsilon}}^\her \mathbf{E}\right) \mathrm{vec} \left(\Delta {\mathbf{\Upsilon}}^\her \right) \right\} \nonumber \\
		& \hspace{.4cm} + \mathrm{tr} \left(\mathbf{E} \widetilde{\mathbf{\Upsilon}} \mathbf{F} \widetilde{\mathbf{\Upsilon}}^\her \right).
	\end{align} 
	\par Based on Lemma (\ref{lemma:bern}), we need to have $\mathrm{vec} \left( \Delta {\mathbf{\Upsilon}}^\her \right) \sim \mathcal{CN} \left( \mathbf{0},\mathit{\mathbf{I}} \right)$. This requirement explains why much of the existing literature assumes $\mathrm{vec} \left( \Delta {\mathbf{\Upsilon}}^\her \right) = \varepsilon \mathbf{i}$, where $\mathbf{i} \sim  \mathcal{CN} \left( \mathbf{0},\mathit{\mathbf{I}} \right)$. To adapt this assumption to our more general approach, we can rewrite $\mathrm{vec} \left( \Delta {\mathbf{\Upsilon}}^\her \right) = \mathbf{T}^{-1} \mathbf{T} \mathrm{vec} \left( \Delta {\mathbf{\Upsilon}}^\her \right)$, where $\mathbf{T} \mathrm{vec} \left( \Delta {\mathbf{\Upsilon}} \right) \sim \mathcal{CN} \left( \mathbf{0},\mathit{\mathbf{I}} \right)$ can be achieved by the Mahalanobis/ZCA, Cholesky, or PCA whitening transformations \cite{Kessy2015OptimalWA}. For the transformation matrix $\mathbf{T}$, we use $\mathbf{T} = \mathbf{\Sigma}^{-1/2}$, which follows the first transformation. Therefore, in accordance with Lemma (\ref{lemma:bern}), we establish
	\begin{subequations} \label{eq:bernCalc}
		\begin{align}
			& \mathbf{x} = \mathbf{T} \mathrm{vec} \left( \Delta {\mathbf{\Upsilon}}^\her \right), \\
			& \mathbf{U} = \mathbf{T}^{-\her} \left(\mathbf{E}^\trp \otimes \mathbf{F} \right) \mathbf{T}^{-1}, \\
			& \mathbf{u} = \mathbf{T}^{-1} \mathrm{vec} \left( \mathbf{F} {\mathbf{\Upsilon}}^\her_n \mathbf{E}\right), \\
			& u = \mathrm{tr} \left(\mathbf{E} {\mathbf{\Upsilon}} \mathbf{F} {\mathbf{\Upsilon}}^\her \right).
		\end{align}
	\end{subequations}
	By applying (\ref{eq:bernConstrs}) to all uncertain constraints in (\ref{eq:decoupledOpt}) and utilizing the parameters from (\ref{eq:bernCalc}), the robust transmission problem can be effectively addressed using the AO, SDR, and Gaussian randomization techniques through the CVX framework. Our robust transmission approach is detailed in Algorithm \ref{alg:robust}.
	\begin{algorithm}                              
		\caption{Empirical DT-based Robust Transmission Design}
		\label{alg:robust}
		\begin{algorithmic}[1]
			\State \textbf{initialization:} Initialize $\mathbf{\Sigma}^{\left(0\right)}$, and set DT channel ($\widetilde{\mathbf{\Upsilon}}$) coherence block $n=1$, 
			\Repeat 
			\If{performance based on $\mathbf{\Sigma}^{\left(n-1\right)}$ not converged}
			\State Obtain the real-world channel ($\mathbf{\Upsilon}$),
			\State Compute $\mathbf{\Sigma}^{\left(n\right)}$ using $\widetilde{\mathbf{\Upsilon}}$ and $\mathbf{\Upsilon}$ based on (\ref{eq:empCov}),
			\Else 
			\State $\mathbf{\Sigma}^{\left(n\right)} \gets \mathbf{\Sigma}^{\left(n-1\right)}$,		
			\EndIf	
			\State Apply (\ref{eq:bernCalc}) to the constraints containing channel approximation errors in the optimization problem (\ref{eq:decoupledOpt}) by following the formulation in (\ref{eq:formulateToBern}),
			\State Initialize $\mathbf{E}^{\left(0\right)}$, and set $t = 1$,
			\While{the optimization problem not converged}
			\State Optimize (\ref{eq:decoupledOpt}) with respect to $\mathbf{F}_k^{\left(t\right)}$, $\forall k \in \left[K\right]$, with given $\mathbf{E}^{\left(t-1\right)}$,
			\State Apply Gaussian randomization,
			\State Optimize (\ref{eq:decoupledOpt}) with respect to $\mathbf{E}^{\left(t\right)}$, with given $\mathbf{F}_k^{\left(t\right)}$,
			\State Apply Gaussian randomization,
			\State $t \gets t+1$,
			\EndWhile
			\State \textbf{output:} Empirical robust transmission design for channel coherence block $n$,
			\State $n \gets n+1$,
			\Until{the zone of a user is changed.}
		\end{algorithmic}
	\end{algorithm}
	\section{Simulation}
	\begin{figure} [t] 
		\centerline{\includegraphics[width=\columnwidth]{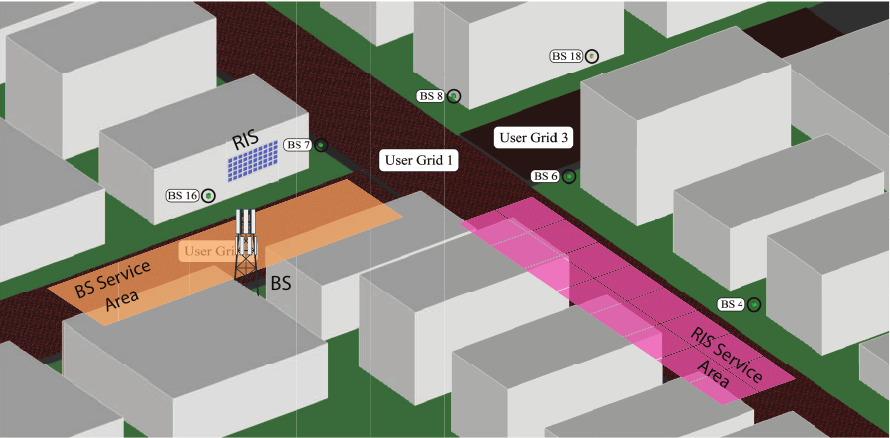}}
		\caption{This figure depicts the geometric arrangement of the simulation scenario, based on the O1 scenario from the DeepMIMO dataset. In this setup, "BS 15" functions as the BS, while "BS 7" serves as the RIS. The highlighted orange and pink regions represent the service area of the BS and RIS, respectively.
		}
		\label{fig:O1}
	\end{figure}
	In this section, we assess the performance of our proposed digital twin-aided framework using the O1 scenario from the DeepMIMO dataset \cite{Alkhateeb2019}, as shown in Fig. \ref{fig:O1}, and compare it with the beam sweeping approach as the baseline. 
	\subsection{Simulation Setup}
	For this evaluation, we designate "BS 15" to function as the base station and "BS 7" to act as the RIS. Both are positioned at a height of $6\text{m}$ and are equipped with $16$ antennas each. The orange grid represents the service area of the BS, while the pink grid indicates the service area of the RIS, which is in a NLoS position relative to the BS, aligning with our system model considerations. The user nodes within these grids are spaced $20\text{cm}$ apart, with each node equipped with a single antenna.
	\subsection{Codebook Generation for Beam Sweeping}
	We develop two distinct codebooks tailored for BS beamforming and RIS reflection, each containing $16$ entries. This results in a combination of $256$ beam pairs. Given that the BS employs a uniform linear array (ULA), its codebook focuses on angular beamforming. In contrast, the RIS's codebook emphasizes spatial beamforming adjustments. Below, we detail the methodology for generating these codebooks.
	\par \textbf{BS Codebook:}
	For the BS codebook generation, we chose 16 angular directions aimed towards the orange grid and computed the optimal beamforming vectors for each direction. One of these directions must align with the RIS to ensure it can effectively serve the user located within the pink grid. This approach does not account for potential interference since it does not incorporate any specific knowledge of the channel characteristics.
	\par \textbf{RIS Codebook:}
	The pink grid was divided into 16 segments for the RIS codebook design, with each segment's optimal RIS phase-shifts being determined based on the angle-of-arrival (AoA) at the user positioned at the center of each segment. The design strategy employed here was conjugate phase beamforming, which optimizes the phase-shifts to maximize signal strength at the user positions.
	\subsection{Imperfect Channel Generation}
	Due to the susceptibility of digital twins to approximation errors from inaccuracies in EM 3D models and ray-tracing parameters, for each user position in the DeepMIMO dataset, we generate channels with $2$ and $10$ strongest paths. We use the channels with $2$ paths as the imperfect DT channels and the channels with $10$ paths as the perfect real-world channels, and consider the difference between them as the DT channel approximation error. To simplify the robust design process, we incorporate these approximation errors only in constraint (\ref{eq:constr4}), which pertains to the user served by the BS. We consider the outage probability $\rho = 0.05$ in (\ref{eq:prob}). The simulations are conducted under the assumption of a perfect $\mathbf{\Sigma}$. However, our numerical results indicate that, despite lacking initial knowledge about the parameters of the channel approximation error distribution, the use of feasible historical real-world channels for $200$ channel coherence blocks allows our empirical robust algorithm to converge to the target performance. After this period, the continued use of costly real-world channels is unnecessary.
	\subsection{Simulation Results}
	\begin{figure} [t]
		\centerline{\includegraphics[width=\columnwidth]{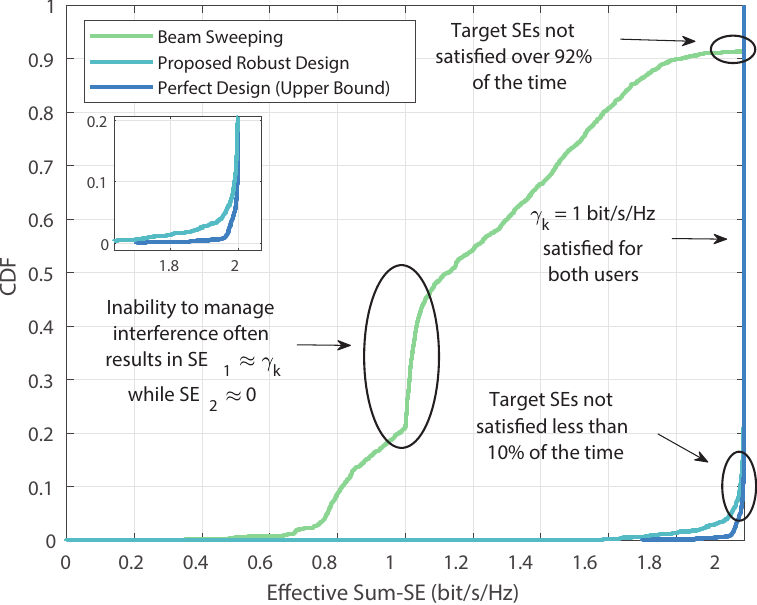}}
		\caption{Cumulative density function (CDF) of effective sum spectral efficiencies (Sum-SE) for a target (minimum required) SE ($\gamma_k = 1$ bit/s/Hz)
		}
		\label{fig:empCDF}
	\end{figure}
	\begin{figure} [t]
		\centerline{\includegraphics[width=\columnwidth]{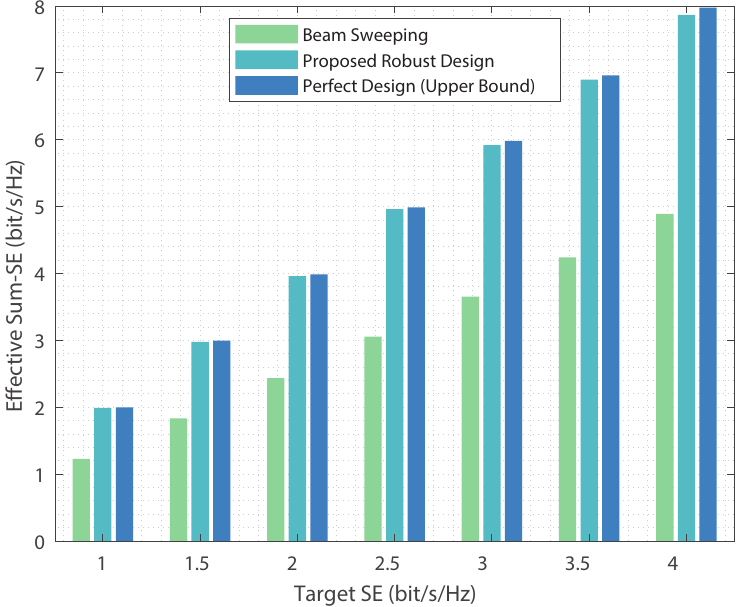}}
		\caption{Effective sum-SE of the users relative to target SEs ($\gamma_k$)
		}
		\label{fig:bar}
	\end{figure}
	Fig. \ref{fig:empCDF} depicts the cumulative density function (CDF) of the effective sum spectral efficiencies achieved in numerical simulations, which employs optimizations based on both perfect and imperfect digital twin channels alongside beam sweeping. Ideally, for a target spectral efficiency, $\gamma_k$, the curves should align with or fall to the right of the line $\text{Sum-SE} = 2\gamma_k$, ensuring that the spectral efficiencies meet the minimum requirements. The outcomes from the perfect design scenario closely match the required lines, with only a few results deviating due to approximations in our optimization methods. The robust design approximates the ideal scenario well, exhibiting less than $10\%$ outage probability. The transmitting power derived from robust optimization at the BS slightly exceeds that from optimizations using perfect digital twin channels. This power level, intended for beam sweeping, leads to an outage probability over $92\%$ and often fails to satisfy the target SEs. The jump in the beam sweeping curve at $\text{Sum-SE}=\gamma_k$ shows that beam sweeping met the target SE for only one user, leaving the other with low SE.
	\par Fig. \ref{fig:bar} shows the effective sum-SEs for different target SEs. The proposed robust design nearly mirrors the performance of the ideal design and meets the required SEs for both users with high probability, while beam sweeping is unable to meet the SE requirements for both users at the same time. 
	\section{Conclusion}
	In this paper, we proposed a novel approach to circumvent the costly channel estimation typically required in RIS-aided wireless networks, and address the shortcomings of more practical methods like beam sweeping that struggle to balance signal power and interference in multi-user settings. We employed digital twins of the physical environments to optimize the network using approximations of real-world channels. To ensure reliable communications, we focused on a primary source of errors in DT channel approximations and developed a DT-specific robust transmission design. This design aims to consistently meet users' needs in terms of effective spectral efficiency while minimizing transmitting power at the BS.

\balance
\bibliographystyle{ieeetr}

\end{document}